\RequirePackage{ifpdf}
\ifpdf 
\documentclass[pdftex]{sigma}
\else
\documentclass{sigma}
\fi

\numberwithin{equation}{section}

\font\sevenrm=cmr7
\newcommand{\J}[1]{J^{\infty} (#1)}
\newcommand{\hp}{\hphantom}
\newcommand{\im}{\mathrm{i}}
\newcommand{\e}{\epsilon}
\newcommand{\R}{\mathcal R^\infty(E_s)}

\newcommand{\hook}
{\mathbin{\raise1.5pt\hbox{\hbox{{\vbox{\hrule height.4pt
width6pt depth0pt}}}\vrule height3pt width.4pt depth0pt}\,}}
\newcommand{\pr}{\operatorname{pr}}
\newcommand{\phiconj}{\overline\phi}
\newcommand\fminus{\hbox{\it F\kern1pt\raise6pt\hbox{\sevenrm +}}}
\newcommand\Wplus{\hbox{$\mathcal W$\kern1pt\raise6pt\hbox{\sevenrm +}}}
\newcommand\hodge{{\mspace{1mu}*}}
\newcommand{\spartial}{\ensuremath \raisebox{0.025cm}{\slash}\hspace{-0.21cm}\partial\/}
\newcommand{\vslash}{\ensuremath \raisebox{0.025cm}{\slash}\hspace{-0.22cm}v\/}

\begin{document}

\renewcommand{\PaperNumber}{004}

\FirstPageHeading

\renewcommand{\thefootnote}{$\star$}

\ShortArticleName{Generalized Symmetries of Massless Free Fields on Minkowski Space}

\ArticleName{Generalized Symmetries of Massless Free Fields\\ on Minkowski Space\footnote{This
paper is a contribution to the Proceedings of the Seventh
International Conference ``Symmetry in Nonlinear Mathematical
Physics'' (June 24--30, 2007, Kyiv, Ukraine). The full collection
is available at
\href{http://www.emis.de/journals/SIGMA/symmetry2007.html}{http://www.emis.de/journals/SIGMA/symmetry2007.html}}}

\Author{Juha POHJANPELTO~$^\dag$ and Stephen C. ANCO~$^\ddag$}

\AuthorNameForHeading{J. Pohjanpelto and S.C. Anco}

\Address{$^\dag$~Department of Mathematics,
Oregon State University,
           Corvallis, Oregon 97331-4605,
           USA}
\EmailD{\href{mailto:juha@math.oregonstate.edu}{juha@math.oregonstate.edu}}
\URLaddressD{\url{http://oregonstate.edu/~pohjanpp/}}

\Address{$^\ddag$~Department of Mathematics,
	Brock University,
                St. Catharines ON L2S 3A1 Canada}
\EmailD{\href{mailto:sanco@brocku.ca}{sanco@brocku.ca}} 
\URLaddressD{\url{http://www.brocku.ca/mathematics/people/anco/}}

\ArticleDates{Received November 01, 2007; Published online January 12, 2008}

\Abstract{A complete and explicit classif\/ication of generalized,
or local,
symmetries of massless free f\/ields of spin $s\geq1/2$
is carried out. Up to equivalence, these are found
to consists of the conformal symmetries and their duals,
new chiral symmetries of order $2s$, and their higher-order
extensions obtained by Lie dif\/ferentiation
with respect to conformal Killing vectors.
In particular, the results yield a complete classif\/ication
of generalized symmetries of the Dirac--Weyl neutrino equation,
Maxwell's equations, and the linearized gravity equations.}


\Classification{58J70; 70S10}

\section{Introduction}
\label{S:intro}
Recent years have seen a growing interest in the study of
the symmetry structure of the main f\/ield equations originating in
mathematical physics.
Generalized, or local, symmetries, which arise as vectors
that are tangent to the solution jet space and preserve the contact
ideal, are important for several reasons. Besides their
original application to the construction of conservation laws,
they play a central role in various methods, in particular
in the classical symmetry reduction \cite{Olver93},
Vessiot's method of group foliation \cite{MaShWi01},
and separation of variables, for f\/inding exact solutions
to systems of partial dif\/ferential equations. Generalized
symmetries also arise in the study of inf\/inite dimensional Hamiltonian systems \cite{Olver93} and are, moreover, connected with B\"acklund
transformations and integrability \cite{Kumei75}.
In fact, the existence of an inf\/inite number of independent generalized symmetries has been proposed as a test for complete integrability of a system of dif\/ferential equations \cite{Mikhailov91}.

For the important examples of
the Einstein gravitational f\/ield equations
and the Yang--Mills f\/ield equations with semi-simple
structure group,
classif\/ications of their symmetry structures~\mbox{\cite{AnTo96, Jp02YM}}
have shown that, besides the obvious gauge symmetries, these
equations essentially admit no generalized symmetries.
In contrast, the linear graviton equations
and the linear Abelian Yang--Mills equations
possess a rich structure of generalized symmetries,
which to-date has yet to be fully determined.

In this paper we present a complete, explicit classif\/ication of
the generalized symmetries for the massless f\/ield equations of
any spin $s=\tfrac{1}{2},1,\tfrac{3}{2},\dots$ on Minkowski
spacetime, formulated in terms of spinor f\/ields.
These equations comprise as special cases
Maxwell's equations, i.e., $U(1)$ Yang--Mills equations for $s=1$,
and graviton equations, i.e., linearized Einstein equations
for $s=2$.
Other f\/ield equations of physical interest which are
included are the Dirac--Weyl, or massless neutrino equation,
and the gravitino equation,
corresponding to the spin values $s=\tfrac{1}{2}$ and $s=\tfrac{3}{2}$, respectively.

There are two main results of the classif\/ication.
First, we obtain spin $s$ generalizations of constant coef\/f\/icient linear
second order symmetries
found some years ago by Fushchich and Nikitin for Maxwell's equations
\cite{FuNi87} and subsequently generalized by Pohjanpelto \cite{Review95}.
The new second order symmetries for Maxwell's equations are especially interesting because,
under duality rotations of the electromagnetic spinor,
they possess odd parity as opposed to the even parity of
the well-known conformal point symmetries.
Consequently, we will refer to the new spin $s$ generalizations
as chiral symmetries.
Furthermore, we show that the spacetime symmetries and chiral symmetries,
together with scalings and duality rotations,
generate the complete enveloping algebra of all generalized symmetries
for the massless spin $s$ f\/ield equations.
In particular, these equations admit no other generalized symmetries apart
from the elementary ones arising from the linearity of the f\/ield equations.
It is also worth noting that, due to the conformal invariance of the massless f\/ield
equations, our results provide as a by-product a complete classif\/ication
of generalized symmetries of spin $s$ f\/ields on any locally
conformally f\/lat spacetime, extending earlier results for the electromagnetic f\/ield obtained by Kalnins et al.~\cite{Kalnins92}.

This classif\/ication is a counterpart to our results
classifying all local conservation laws
for the massless spin $s$ f\/ield equations \cite{sAjP1,sAjP2,sAjP3}.
We emphasize, however, that there is no
immediate Noether correspondence between
conservation laws and symmetries in our situation,
as the formulation of the massless spin $s$ f\/ield equations
in terms of spinor f\/ields does not admit a local Lagrangian.

Our paper is organized as follows. First, in Section~\ref{S:preliminaries} we cover some background material on symmetries of dif\/ferential equations and on spinorial formalism, including a factorization property of Killing spinors on Minkowski space that is pivotal in the symmetry analysis carried out in this paper. Then in Section~\ref{S:MainResults} we state and prove our classif\/ication theorem for generalized symmetries of arbitrary order for the massless f\/ield equations. These, in particular, include novel chiral symmetries of order $2s$ for
spin $s=\tfrac{1}{2},1,\tfrac{3}{2},\dots$ f\/ields.
As applications, in Section~\ref{S:MaxwellSymmetries}
we transcribe our main result in the spin $s=1$ case into tensorial form to derive a complete classif\/ication of generalized symmetries for the vacuum Maxwell's equations, and, f\/inally, in Section~\ref{S:Dirac}, we employ the methods of Section~\ref{S:MainResults} to carry out a full symmetry analysis of the Weyl system, or the massless Dirac equation,
on Minkowski space. Our results for the Weyl system complement
those found in \cite{BeKr04, DuLiVi88, Niki91}, and, in particular, provide a classif\/ication of symmetries of arbitrarily high order for the massless neutrino equations.

\section{Preliminaries}
\label{S:preliminaries}

Let $\mathbf M$
be Minkowski space with coordinates
$x^i$, $0\leq i\leq3$, and let $E_s$,
$s=\tfrac{1}{2}$, $1$, $\tfrac{3}{2}$, \dots,
stand for the coordinate bundle
\begin{displaymath}
\pi:E_s = \{(x^i,\phi_{A_1A_2\cdots A_{2s}})\}\to\{(x^i)\},
\end{displaymath}
where $\phi_{A_1A_2\cdots A_{2s}}$ is a type $(2s,0)$
spinor. We denote the $k$th order jet bundle of
local sections of
$E_s$ by $J^k(E_s)$, $0\leq k\leq\infty$.
Recall that the inf\/inite
jet bundle $\J{E_s}$ is the coordinate space
\begin{displaymath}
\J{E_s} = \{(x^i,\phi_{A_1A_2\cdots A_{2s}},
\phi_{A_1A_2\cdots A_{2s},j_1},
\dots,\phi_{A_1A_2\cdots A_{2s},j_1j_2\cdots j_p},\dots
)\},
\end{displaymath}
where $\phi_{A_1A_2\cdots A_{2s},j_1j_2\cdots j_p}$
stands for the $p$th order derivative variables.
As is customary, we write
\begin{displaymath}
\phi_{A_1A_2\cdots A_{2s}, B_1B_2\cdots B_p}^
{\hp{A_1A_2\cdots A_{2s},}B_1'B_2'\cdots B_p'}=
\sigma^{j_1}{}_{B_1}^{B_1'}
\sigma^{j_2}{}_{B_2}^{B_2'}\cdots
\sigma^{j_p}{}_{B_p}^{B_p'}
\phi_{A_1A_2\cdots A_{2s},j_1j_2\cdots j_p},
\end{displaymath}
where $\sigma^{j}{}_{BB'}$, $0\leq j\leq3$, are, up to
a constant factor, the identity matrix and the Pauli
spin matrices.
We also write
\begin{displaymath}
\overline\phi_{A_1'A_2'\cdots A_{2s}', B_1'\cdots B_p'}^
{\hp{A_1'A_2'\cdots A_{2s}',}B_1\cdots B_p}=
\overline{\phi_{A_1A_2\cdots A_{2s}, B_1\cdots B_p}^
{\hp{A_1A_2\cdots A_{2s},}B_1'\cdots B_p'}}\,,
\end{displaymath}
where the bar stands for complex conjugation.
Here and in the sequel we employ the Einstein summation convention in both the space-time and spinorial indices, and we lower and raise spinorial
indices using the spinor metric $\e_{AB}$ and its inverse
$\e^{AB}$; see \cite{Penrose84} for further details.

In order to streamline our notation, we will employ boldface capital letters to designate spinorial multi-indices.
Thus, for example, we will write
\begin{displaymath}
\phi_{\mathbf A_{2s}, \mathbf B_p}^
{\hp{\mathbf A_{2s},}\mathbf B_p'}=
\phi_{A_1A_2\cdots A_{2s}, B_1B_2\cdots B_p}^
{\hp{A_1A_2\cdots A_{2s},}B_1'B_2'\cdots B_p'},
\end{displaymath}
and we will combine multi-indices by
the rule $\mathbf B_p\mathbf C_q=
(B_1B_2\cdots B_p C_1C_2\cdots C_q)$.

We let
\begin{displaymath}
\partial_{CC'} = \sigma_{CC'}^i\partial/\partial x^i
\end{displaymath}
denote the spinor representative of
the coordinate derivative $\partial/\partial x^i$.
Moreover, we def\/ine partial derivative operators
$\partial^{\mathbf A_{2s},\mathbf B_p}_
{\phi\hp{A_{2s}}\mathbf B_p'}$ by
\begin{align}
\partial^{\mathbf A_{2s},\mathbf B_p}_
{\phi\hp{A_{2s}}\mathbf B_p'}
\phi_{\mathbf C_{2s}, \mathbf D_r}^
{\hp{\mathbf C_{2s},}\mathbf D_r'}
&= \begin{cases}
\e_{(C_1}{}^{A1}\cdots \e_{C_{2s})}{}^{A_{2s}}
\e_{(D_1}{}^{B_1}\cdots \e_{D_p)}{}^{B_p}
\e_{(B_1'}{}^{D_1'}\cdots \e_{B_p')}{}^{D_p'},
&\text{if \ \ $p=r$},\\ 0,&\text{if \ \ $p\neq r$,}
\end{cases}\nonumber\\
\partial^{\mathbf A_{2s},\mathbf B_p}_
{\phi\hp{A_{2s}}\mathbf B_p'}
\overline\phi_{\mathbf C_{2s}', \mathbf D_r'}^
{\hp{\mathbf C_{2s}',}\mathbf D_r}&=0,\nonumber
\end{align}
and write
\begin{displaymath}
\overline\partial^{\mathbf A_{2s}',\mathbf B_p'}_
{\phi\hp{A_{2s}}\mathbf B_p}=
\overline{\partial^{\mathbf A_{2s},\mathbf B_p}_
{\phi\hp{A_{2s}}\mathbf B_p'}}\;.
\end{displaymath}
Here, in accordance with the standard spinorial notation, we have
written $\e_C{}^A$ for the Kronecker delta and we use round
brackets to indicate symmetrization in the enclosed indices.

A generalized vector f\/ield $X$ on $E_s$
in spinor form is a vector f\/ield
\begin{equation}
\label{E:GeneralizedVectorField}
X = P^{CC'}\partial_{CC'}+
Q_{\mathbf A_{2s}}\partial^{\mathbf A_{2s}}_\phi+
\overline Q_{\mathbf A_{2s}'}
\overline\partial^{\mathbf A_{2s}'}_\phi,
\end{equation}
where the coef\/f\/icients
$P^{CC'}=P^{CC'}(x^j, \phi^{[p]})$,
$Q_{\mathbf A_{2s}}=Q_{\mathbf A_{2s}}(x^j,\phi^{[p]})$
are spinor valued functions in $x^j$
and the derivative variables
$\phi_{\mathbf A_{2s}, \mathbf B_q}^
{\hp{\mathbf A_{2s},}\mathbf B_q'}$
up to some f\/inite order $p$.
An evolutionary vector f\/ield~$Y$, in turn, is a
generalized vector f\/ield of the form
\begin{displaymath}
Y = Q_{\mathbf A_{2s}}\partial^{\mathbf A_{2s}}_{\phi}+
\overline Q_{\mathbf A_{2s}'}
\overline\partial^{\mathbf A_{2s}'}_{\phi},
\end{displaymath}
 where $Q_{\mathbf A_{2s}}$ is called the characteristic of $Y$.

Let
\begin{equation*}
D_{C}^{C'} = \partial_{C}^{C'} +
\sum_{p\geq0}\big(\phi_{\mathbf A_{2s},\mathbf B_pC}^
{\hp{\mathbf A_{2s},}\mathbf B_p'C'}
\partial^{\mathbf A_{2s},\mathbf B_p}_
{\phi\hp{A_{2s}}\mathbf B_p'}+
\overline\phi_{\mathbf A_{2s}',\mathbf B_p'C}^
{\hp{\mathbf A_{2s}',}\mathbf B_pC'}
\overline\partial^{\mathbf A_{2s}',\mathbf B_p'}_
{\phi\hp{A_{2s}}\mathbf B_p}\big)
\end{equation*}
stand for the spinor representative of the
standard total derivative operator, which, as is
easily verif\/ied, satisf\/ies the commutation formula
\begin{equation}
\label{E:CommutationFormula}
[\partial^{\mathbf A_{2s},\mathbf B_p}_
{\phi\hp{A_{2s}}\mathbf B_p'},D_C^{C'}]
=\e_{(B_p'|}{}^{C'}\e_{C}{}^{(B_p|}
\partial^{\mathbf A_{2s},|\mathbf B_{p-1})}_
{\phi\hp{A_{2s}}|\mathbf B_{p-1}')},\qquad p\geq1.
\end{equation}
The inf\/inite prolongation $\pr X$ of $X$
in \eqref{E:GeneralizedVectorField} to a
vector f\/ield on $\J{E_s}$ is given by
\begin{equation*}
\pr X = P^{CC'}D_{CC'}
+\sum_{p\geq 0}\bigl((D_{B_1}^{B_1'}\cdots
D_{B_p}^{B_p'}R_{\mathbf A_{2s}})
\partial^{\mathbf A_{2s},\mathbf B_p}_
{\phi\hp{A_{2s}}\mathbf B_p'}+
(D_{B_1}^{B_1'}\cdots
D_{B_p}^{B_p'}\overline R_{\mathbf A_{2s}'})
\overline\partial^{\mathbf A_{2s}',\mathbf B_p}_
{\phi\hp{A_{2s}'}\mathbf B_p'}\bigr),
\end{equation*}
where $R_{\mathbf A_{2s}}$ is the characteristic of the evolutionary form
\begin{displaymath}
X_{\text{ev}} = (Q_{\mathbf A_{2s}}-
P^{CC'} \phi_{\mathbf A_{2s},CC'})
\partial^{\mathbf A_{2s}}_\phi+
(\overline Q_{\mathbf A_{2s}'}-
P^{CC'} \overline\phi_{\mathbf A_{2s}',CC'})
\overline\partial^{\mathbf A_{2s}'}_\phi
\end{displaymath}
of $X$.

The massless free f\/ield equation of spin
$s$ and its dif\/ferential consequences
\begin{equation}
\phi_{\mathbf A_{2s},A'\hskip5pt\mathbf B_{p}}^
{\hp{\mathbf A_{2s},}A_{2s}
\mathbf B_p'} = 0,
\qquad p\geq0,
\label{E:MasslessEquations}
\end{equation}
determine the inf\/initely prolonged solution
manifold $\R\subset J^\infty(E_s)$ of the equations.
According to \cite{Penrose84}, the
symmetrized derivative variables
\begin{equation*}
\phi_{\mathbf A_{2s}\mathbf B_p}^{\hp{\mathbf A_{2s}}\mathbf B_p'}=
\phi_{(\mathbf A_{2s},\mathbf B_p)}^{\hp{(\mathbf A_{2s},}\mathbf B_p'},
\qquad p\geq 0,
\end{equation*}
known as Penrose's exact sets of f\/ields, together with the independent variables $x^i$ provide
coordinates for $\R$. Moreover, as is easily verif\/ied,
the unsymmetrized and symmetrized variables
$\phi_{\mathbf A_{2s},\mathbf B_p}^{\hp{A_{2s},}\mathbf B_p'}$,
$\phi_{\mathbf A_{2s}\mathbf B_p}^{\hp{A_{2s}}\mathbf B_p'}$
agree on $\R$.

A generalized, or local, symmetry of massless free f\/ields
is a generalized vector f\/ield $X$ satisfying
\begin{equation}
\label{E:DeterminingEquations}
\pr X\phi_{\mathbf A_{2s}, A'}^
{\hp{\mathbf A_{2s},}A_{2s}} =0\qquad\mbox{on \ \ $\R$}.
\end{equation}
Note that any generalized vector f\/ield of the form
\begin{displaymath}
T_P = P^{CC'}(\partial_{CC'}+
\phi_{\mathbf A_{2s}, CC'}\partial^{\mathbf A_{2s}}_\phi+
\overline\phi_{\mathbf A_{2s}', CC'}
\overline\partial^{\mathbf A_{2s}'}_\phi)
\end{displaymath}
with the prolongation
\begin{displaymath}
\pr T_P = P^{CC'}D_{CC'}
\end{displaymath}
automatically
satisf\/ies the determining equations
\eqref{E:DeterminingEquations} for
a~symmetry.
Hence we will call a symmetry trivial
if its prolongation agrees with a total vector
f\/ield $\pr T_P = P^{CC'}D_{CC'}$
on~$\R$ and we call two
symmetries equivalent if their dif\/ference
is a trivial symmetry. See, e.g., \cite{book, Olver93}
for further details and background material on
generalized symmetries.

In this paper we explicitly
classify all equivalence classes of generalized
symmetries of massless free f\/ields of spin
$s=\tfrac{1}{2},1,\tfrac{3}{2},\dots$ on Minkowski space.
By the above, in our classif\/ication we only need to
consider symmetries in evolutionary form, and
for such a vector f\/ield
\begin{displaymath}
Y = Q_{\mathbf A_{2s}}\partial^{\mathbf A_{2s}}_\phi+
\overline Q_{\mathbf A_{2s}'}
\overline\partial^{\mathbf A_{2s}'}_\phi,
\end{displaymath}
the determining equations \eqref{E:DeterminingEquations}
for the characteristic $Q_{\mathbf A_{2s}}$ become
\begin{equation}
\label{E:EvolutionarySymmetry}
D_{A'}^{A_{2s}}Q_{\mathbf A_{2s}} = 0\qquad
\mbox{on \ \ $\R$}.
\end{equation}
Moreover, after replacing $X$ by an
equivalent symmetry, we can always assume
that the components $Q_{\mathbf A_{2s}}$
are functions of only the independent variables
$x^i$ and the symmetrized derivative
variables
$\phi_{\mathbf B_{2s+q}}^
{\mathbf B_q'}$, $0\leq q\leq p$,
for some $p$.

Recall that a vector f\/ield $\xi=\xi^i(x^j)\partial_i$
on $\mathbf M$ is conformal Killing provided
that
\begin{equation}
\label{E:KillingEquation}
\partial_{(i}\xi_{j)}=k\eta_{ij}
\end{equation} for some function
$k=k(x^i)$, where $\eta_{ij}$ stands
for the Minkowski metric.
As can be verif\/ied by a direct computation,
a conformal Killing vector
f\/ield $\xi$ gives rise to the symmetry
\begin{equation}
\label{E:ConformalSymmetry}
\mathcal Z[\xi]=
\mathcal Z_{\mathbf A_{2s}}[\xi]\partial_{\phi}^{\mathbf A_{2s}}+
\overline{\mathcal Z}_{\mathbf A_{2s}'}[\xi]
\overline\partial_{\phi}^{\mathbf A_{2s}'}
\end{equation}
of massless free
f\/ields of spin $s$, with the characteristic
\begin{equation*}
\mathcal Z_{\mathbf A_{2s}}[\xi] =
\xi^{CC'}\phi_{\mathbf A_{2s}CC'}+
s\partial_{C'(A_{2s}}\xi^{CC'}\phi_{\mathbf A_{2s-1})C}+
\dfrac{1-s}{4}(\partial_{CC'}\xi^{CC'})\phi_{\mathbf A_{2s}},
\end{equation*}
which agrees with the conformally weighted Lie derivative
\begin{displaymath}
\mathcal L^{(-1)}_\xi\phi_{\mathbf A_{2s}}=
\mathcal L_\xi\phi_{\mathbf A_{2s}} +
\dfrac{1}{4}(\partial_{CC'}\xi^{CC'})\phi_{\mathbf A_{2s}}
\end{displaymath}
of the spinor f\/ield $\phi_{\mathbf A_{2s}}$; see
\cite{sAjP2, Penrose84}.

In the course of the present symmetry classif\/ication
we will repeatedly use the fact that, on account
of the linearity of the massless free f\/ield equations,
the componentwise derivative
\begin{displaymath}
\pr \mathcal Z[\xi]Y
\end{displaymath}
of an evolutionary symmetry $Y$ with respect
to the prolongation of the vector f\/ield
$\mathcal Z[\xi]$
is again a symmetry of the equations;
see, e.g.,~\cite{Review95}.

In spinor form the conformal Killing vector
equation \eqref{E:KillingEquation} becomes
\begin{equation}
\label{E:SpinorFormOfKillingVector}
\partial_{(B}^{(B'}\xi_{C)}^{C')}=0.
\end{equation}
An obvious generalization of equations
\eqref{E:SpinorFormOfKillingVector} to spinor
f\/ields $\kappa^{\mathbf A'_l}_{\mathbf A_k} =
\kappa^{\mathbf A'_l}_{\mathbf A_k}(x^{CC'})$
of type $(k,l)$ is
\begin{equation}
\label{E:KillingSpinorEquation}
\partial^{(A_{l+1}'}_{(A_{k+1}}
\kappa^{\mathbf A'_l)}_{\mathbf A_k)} = 0,
\end{equation}
and symmetric spinor f\/ields
$\kappa^{\mathbf A'_l}_{\mathbf A_k} =
\kappa^{(\mathbf A'_l)}_{(\mathbf A_k)}(x^{CC'})$
satisfying these equations
are called Killing spinors of type $(k,l)$.
Thus, in particular, a type $(1,1)$ Killing spinor
$\kappa^{A'}_{A}$
corresponds to a complex conformal Killing vector.
The following Lemma, which is a special case of
the well-known factorization property of Killing spinors
on Minkowski space, is pivotal in our classif\/ication
of symmetries of massless free f\/ields. For more
details, see \cite{Penrose84}.
\begin{lemma}
\label{L:KillingSpinor}
Let $\xi^{\mathbf A'_k}_{\mathbf A_k}$,
$\kappa^{\mathbf A'_{k+2s}}_{\mathbf A_{k}}$
be Killing spinors of type $(k,k)$ and $(k,k+2s)$.
Then $\xi^{\mathbf A'_k}_{\mathbf A_k}$
can be expressed as a sum of symmetrized
products of $k$ Killing spinors of type $(1,1)$,
and $\kappa^{\mathbf A'_{k+2s}}_{\mathbf A_{k}}$
can be expressed as a sum of symmetrized products
of Killing spinors of type $(0,2s)$
and $k$ Killing spinors of type $(1,1)$.
The dimensions of the complex vector spaces of
Killing spinors of type~$(k,k)$ and $(k,k+2s)$
are
\begin{align}
&(k+1)^2(k+2)^2(2k+3)/12\qquad\quad\text{and}\nonumber\\
&(k+1)(k+2)(k+2s+1)(k+2s+2)(2k+2s+3)/12,\nonumber
\end{align}
respectively.
\end{lemma}

\section{Main results}
\label{S:MainResults}

Let $\xi$, $\zeta_1,\dots, \zeta_p$ be real
conformal Killing vectors and let
$\pi^{\mathbf A_{4s}'}$ be a type
$(0,4s)$ Killing spinor. Let
$\mathcal Z[\xi]$ be the symmetry associated
with $\xi$ as in \eqref{E:ConformalSymmetry}
and def\/ine $\mathcal Z[\im\xi]$ by
\begin{equation}
\label{E:DualOfConformalSymmetry}
\mathcal Z[\im\xi]=
\im\mathcal Z_{\mathbf A_{2s}}[\xi]\partial_{\phi}^{\mathbf A_{2s}}-
\im\overline{\mathcal Z}_{\mathbf A_{2s}'}[\xi]
\overline\partial_{\phi}^{\mathbf A_{2s}'}.
\end{equation}
Furthermore, let
\begin{equation}
\label{E:ChiralSymmetry}
\mathcal W[\pi]=
\mathcal W_{\mathbf A_{2s}}[\pi]
\partial_\phi^{\mathbf A_{2s}}+
\overline{\mathcal W}_{\mathbf A_{2s}'}[\pi]
\overline\partial_\phi^{\mathbf A_{2s}'}
\end{equation}
be an evolutionary vector
f\/ield with components
\begin{equation}
\label{E:ChiralSymmetryComponents}
\mathcal W_{\mathbf A_{2s}}[\pi]= \sum_{p=0}^{2s}
\text{c}_{2s,p} \partial_{B_1'(A_{2s-p+1}|}
\partial_{B_2'|A_{2s-p+2}|}\cdots
_|\partial_{B_p'|A_{2s}}\pi^{\mathbf B_p'\mathbf  C_{4s-p}'}
\overline\phi_{\mathbf A_{2s-p})\mathbf C_{4s-p}'},
\end{equation}
where the coef\/f\/icients $\text{c}_{2s,p}$ are
given by
\begin{equation}\label{E:ChiralCoeffs}
\text{c}_{2s,p} = \dfrac{4s-p+1}{4s+1}\binom{2s}{p},
\qquad 0\leq p\leq 2s.
\end{equation}
Moreover, write
\begin{align}
\mathcal Z[\xi;\zeta_1,\dots,\zeta_p] &=
\pr \mathcal Z[\zeta_1]\cdots\pr \mathcal Z[\zeta_p]
\mathcal Z[\xi],
\label{E:HigherConformalSymmetry}\\
\mathcal Z[\im \xi;\zeta_1,\dots,\zeta_p] &=
\pr \mathcal Z[\zeta_1]\cdots\pr \mathcal Z[\zeta_p]
\mathcal Z[\im\xi],
\label{E:DualOfHigherConformalSymmetry}\\
\mathcal W[\pi;\zeta_1,\dots,\zeta_q] &=
\pr \mathcal Z[\zeta_1]\cdots\pr \mathcal Z[\zeta_q]
\mathcal W[\pi],
\label{E:HigherChiralSymmetry}
\end{align}
for the repeated componentwise derivatives of the
vector f\/ields \eqref{E:ConformalSymmetry}, \eqref{E:DualOfConformalSymmetry}, \eqref{E:ChiralSymmetry}
with respect to conformal symmetries.
\begin{proposition}
\label{P:LeadingOrderTerms}
Let $\xi$, $\zeta_1$,\dots,$\zeta_p$
be conformal Killing vectors and
let $\pi_{\mathbf A_{4s}}$ be a
Killing spinor of type $(0,4s)$.
Then the evolutionary vector fields
\begin{equation}\label{E:MasslessSymmetries}
\mathcal Z[\xi;\zeta_1,\dots,\zeta_p],\qquad
\mathcal Z[\im\xi;\zeta_1,\dots,\zeta_p],\qquad
\mathcal W[\pi;\zeta_1,\dots,\zeta_q],
\qquad p,q\geq 0,
\end{equation}
are symmetries of the massless free field
equations of spin $s$ of order $p+1$ and $q+2s$,
respectively. Moreover, when restricted to the
solution manifold $\R$, the leading order
terms in the components
$\mathcal Z_{\mathbf A_{2s}}[\xi;\zeta_1,\dots,\zeta_p]$,
$\mathcal Z_{\mathbf A_{2s}}[\im\xi;\zeta_1,\dots,\zeta_p]$,
$\mathcal W_{\mathbf A_{2s}}[\pi;\zeta_1,\dots,\zeta_q]$
of the symmetries \eqref{E:MasslessSymmetries} reduce to
\begin{align}
&(-1)^{p+1}\xi^{(C_1}_{(C_1'}\zeta_1{}^{C_2}_{C_2'}
\cdots\zeta_p{}^{C_{p+1})}_{C_{p+1}')}
\phi_{\mathbf A_{2s}\mathbf C_{p+1}}^
{\hp{\mathbf A_{2s}}\mathbf C_{p+1}'},
\nonumber\\
&(-1)^{p+1}\im\xi^{(C_1}_{(C_1'}\zeta_1{}^{C_2}_{C_2'}
\cdots\zeta_p{}^{C_{p+1})}_{C_{p+1}')}
\phi_{\mathbf A_{2s}\mathbf C_{p+1}}^
{\hp{\mathbf A_{2s}}\mathbf C_{p+1}'},
\nonumber\\
&(-1)^q\zeta_1{}_{(C_1'}^{(C_1}\zeta_2{}_{C_2'}^{C_2}
\cdots \zeta_q{}_{C_q'}^{C_q)}\pi_{\mathbf B_{4s}')}
\phiconj_{\mathbf A_{2s}\mathbf C_q}^
{\hp{\mathbf A_{2s}}\mathbf C_q'\mathbf B_{4s}'},
\nonumber
\end{align}
respectively.
\end{proposition}

\begin{proof} We only need to show that
$\mathcal W[\pi]$ in \eqref{E:ChiralSymmetry},
\eqref{E:ChiralSymmetryComponents}
satisf\/ies the symmetry equations
\eqref{E:DeterminingEquations}.
First note that due to the Killing spinor equations
\eqref{E:KillingSpinorEquation} we have
that
\begin{equation}
\label{E:DlambertOnPi}
\partial_{CC'}\partial^{CD'}\pi^{\mathbf B_{4s}'} = 0,
\end{equation}
and, consequently,
\begin{equation}
\label{E:CommuteDerivativesOfPi}
\partial_{CC'}\partial_{DD'}\pi^{\mathbf B_{4s}'} =
\partial_{CD'}\partial_{DC'}\pi^{\mathbf B_{4s}'}.
\end{equation}

Write
\begin{align}
\label{E:DefinitionOfPi}
\Pi_{p,\mathbf A_{2s-1}A'}^{1} &=
\partial^{A_{2s}}_{A'}\partial_{B_1'(A_{2s-p+1}|}
\partial_{B_2'|A_{2s-p+2}|}\cdots _|
\partial_{B_p'|A_{2s}}\pi^{\mathbf B_{p}'\mathbf C_{4s-p}'}
\overline\phi_{\mathbf A_{2s-p})\mathbf C_{4s-p}'},
\nonumber\\
\Pi_{p,\mathbf A_{2s-1}A'}^{2} &=
\partial_{B_1'(A_{2s-p+1}|}\partial_{B_2'|A_{2s-p+2}|}\cdots _|
\partial_{B_p'|A_{2s}}\pi^{\mathbf B_{p}'\mathbf C_{4s-p}'}
\overline\phi_{\mathbf A_{2s-p})\mathbf C_{4s-p}'A'}^{A_{2s}},
\nonumber
\end{align}
$0\leq p\leq 2s$, so that $\mathcal W[\pi]$ is a symmetry of
massless free f\/ield equations provided that
\begin{equation}
\label{E:SymmetryEquationForPi}
\sum_{p=0}^{2s}\text{c}_{2s,p}(\Pi_{p,\mathbf A_{2s-1}A'}^{1}+
\Pi_{p,\mathbf A_{2s-1}A'}^{2}) =0.
\end{equation}

We compute
\begin{align}
\label{E:Pi1Reduced}
&\Pi_{p,\mathbf A_{2s-1}A'}^{1}\nonumber\\
&=\dfrac{4s}{4s+1}\partial_{B_1'(A_{2s-p+1}|}
\partial_{B_2'|A_{2s-p+2}|}\cdots _|
\partial_{B_p'|A_{2s}}(\e_{A'}{}^{(B_{p}'}\partial_{|D'|}^{|A_{2s}|}
\pi^{\mathbf B_{p-1}'\mathbf C_{4s-p}')D'})
\overline\phi_{\mathbf A_{2s-p})\mathbf C_{4s-p}'}\nonumber\\
&=\dfrac{p}{4s+1}\partial_{A'(A_{2s-p+1}|}
\partial_{B_1'|A_{2s-p+2}|}\cdots _|
\partial_{B_{p-1}'|A_{2s}}\partial_{|D'|}^{A_{2s}}
\pi^{\mathbf B_{p-1}'\mathbf C_{4s-p}'D'}
\overline\phi_{\mathbf A_{2s-p})\mathbf C_{4s-p}'}\nonumber\\
&\qquad+\dfrac{4s-p}{4s+1}\partial_{B_1'(A_{2s-p+1}|}
\partial_{B_2'|A_{2s-p+2}|}\cdots _|
\partial_{B_p'|A_{2s}}\partial_{|D'|}^{A_{2s}}
\pi^{\mathbf B_p'\mathbf C_{4s-p-1}'D'}
\overline\phi_{\mathbf A_{2s-p})\mathbf C_{4s-p-1}'A'}\nonumber\\
&=\dfrac{p}{4s+1}\Pi_{p,\mathbf A_{2s-1}A'}^{1}+\dfrac{4s-p}{4s+1}\dfrac{2s-p}{2s}\\
&\qquad\times\partial_{B_1'(A_{2s-p}|}\partial_{B_2'|A_{2s-p+1}|}\cdots _|
\partial_{B_p'|A_{2s-1}}\partial_{|D'|}^{B}\pi^{\mathbf B_p'\mathbf C_{4s-p-1}'D'}
\overline\phi_{\mathbf A_{2s-p-1})\mathbf C_{4s-p-1}'A'B},\nonumber
\end{align}
where we used \eqref{E:DlambertOnPi} and
\eqref{E:CommuteDerivativesOfPi}.
On the other hand,
\begin{equation}
\label{E:Pi2Reduced}
\Pi_{p,\mathbf A_{2s-1}A'}^{2}=
\dfrac{p}{2s}\partial_{D'B}
\partial_{B_1'(A_{2s-p+1}|}\partial_{B_2'|A_{2s-p+2}|}\cdots _|
\partial_{B_p'|A_{2s-1}}\pi^{\mathbf B_{p}'\mathbf C_{4s-p}'D'}
\overline\phi_{\mathbf A_{2s-p})\mathbf C_{4s-p}'A'}^{B}.
\end{equation}

Now it follows from \eqref{E:Pi1Reduced},
\eqref{E:Pi2Reduced} that
\begin{equation*}
\Pi_{p,\mathbf A_{2s-1}A'}^{1} =
-\dfrac{(4s-p)(2s-p)}{(4s-p+1)(p+1)}
\Pi_{p+1,\mathbf A_{2s-1}A'}^{2}.
\end{equation*}
Clearly
\begin{equation*}
\Pi_{2s,\mathbf A_{2s-1}A'}^{1}=0,\qquad
\Pi_{0,\mathbf A_{2s-1}A'}^{2}=0.
\end{equation*}
Consequently, by virtue of \eqref{E:ChiralCoeffs},
equation \eqref{E:SymmetryEquationForPi}
holds and hence $\mathcal W[\pi]$ is a symmetry
of the massless free f\/ield equations.
\end{proof}

The massless free f\/ield equations of spin $s$ also
admit the obvious scaling symmetry $\mathcal S$,
its dual symmetry $\widetilde{\mathcal S}$, and the elementary
symmetries $\mathcal E[\varphi]$ given by
\begin{equation}
\label{E:ScalingSymmetries}
\mathcal S = \phi_{\mathbf A_{2s}}\partial^{\mathbf A_{2s}}+
\overline\phi_{\mathbf A_{2s}'}\overline\partial^{\mathbf A_{2s}'},\qquad
\widetilde{\mathcal S} = \mbox{\rm i}\phi_{\mathbf A_{2s}}
\partial^{\mathbf A_{2s}}-\mbox{\rm i}\overline
\phi_{\mathbf A_{2s}'}\overline\partial^{\mathbf A_{2s}'},
\end{equation}
and
\begin{equation}
\label{E:TrivialSymmetries}
\mathcal E[\varphi] =
\varphi_{\mathbf A_{2s}}(x^i)\partial^{\mathbf A_{2s}}+
\overline\varphi_{\mathbf A_{2s}'}(x^i)\overline\partial^{\mathbf A_{2s}'},
\end{equation}
where $\varphi_{\mathbf A_{2s}} = \varphi_{\mathbf A_{2s}}(x^i)$  is any solution of \eqref{E:MasslessEquations}.

\begin{theorem}
\label{T:MainTheorem}
Let ${\mathcal Q}$
be a generalized symmetry
of the massless free field equations of spin
$s=\tfrac{1}{2},1,\tfrac{3}{2},\dots$.
If the evolutionary form of $\mathcal Q$ is of
order $r$, then ${\mathcal Q}$ is
equivalent to a symmetry $\widehat{\mathcal Q}$ of
order at most $r$
which can be written as
\begin{equation}
\label{E:RemoveTrivialSymmetry}
\widehat{\mathcal Q} = \mathcal V+\mathcal E[\varphi],
\end{equation}
where $\varphi_{\mathbf A_{2s}} = \varphi_{\mathbf A_{2s}}(x^i)$
is a solution of the massless free field
equations of spin $s$, and where $\mathcal V$
is equivalent to a spinorial symmetry
that is a linear combination of the symmetries
\begin{equation}
\mathcal S,\quad
\widetilde{\mathcal S},
\quad\mathcal Z[\xi;\zeta_1,\dots,\zeta_p],\quad
\mathcal Z[\im\xi;\zeta_1,\dots,\zeta_p],\quad
\mathcal W[\pi;\zeta_1,\dots,\zeta_q]
\label{E:NonTrivialSymmetries}
\end{equation}
with $p\leq r-1$, $q\leq r-2s$.

In particular, the dimension $d_r$ of the
real vector space of equivalence classes
of spinorial symmetries of order at most
$r$ spanned by the symmetries
\eqref{E:NonTrivialSymmetries} is
\[
d_r  =
(r+1)^2(r+2)^2(r+3)^2/18,\qquad\text{if \ \ $r<2s$,}\nonumber\\
\]
and
\begin{align*}
d_r  = {}& (r+1)^2(r+2)^2(r+3)^2/18\\ 
   & {} +((r+1)^2-4s^2)((r+2)^2-4s^2)((r+3)^2-4s^2)/18,\qquad
\text{if \ \ $r\geq 2s$.}\nonumber
\end{align*}
\end{theorem}

The above result was originally announced without proof
in \cite{sAjP3}.

\begin{proof} Without loss of generality,
we can assume that the components
$\mathcal Q_{\mathcal A_{2s}}$ of
$\mathcal Q$
are functions of the independent variables
$x^i$ and the symmetrized derivative variables
$\phi_{\mathbf A_{2s+p}}^{\mathbf A_{p}'}$,
$0\leq p\leq r$. Consequently,
\begin{equation}
\label{E:MaySymmetrizeDerivatives}
\partial^{\hp{\phi}\mathbf B_{2s},\mathbf C_{p}}_
{\phi\hp{\mathbf B_{2s},}\mathbf C_p'}\mathcal Q_{\mathbf A_{2s}}=
\partial^{\hp{\phi}(\mathbf B_{2s},\mathbf C_p)}_
{\phi\hp{(\mathbf B_{2s},}\mathbf C_p'}\mathcal Q_{\mathbf A_{2s}},\quad
\overline\partial^{\hp{\phi}\mathbf B_{2s}',\mathbf C_{p}'}_
{\phi\hp{\mathbf B_{2s}',}\mathbf C_p}\mathcal Q_{\mathbf A_{2s}}=
\overline\partial^{\hp{\phi}(\mathbf B_{2s}',\mathbf C_p')}_
{\phi\hp{(\mathbf B_{2s},}\mathbf C_p}\mathcal Q_{\mathbf A_{2s}}.
\end{equation}

It follows from the determining equations
for spinorial symmetries that
\begin{equation}
\label{E:SymmetrizedDerivatives}
\partial^{\hp{\phi}(\mathbf B_{2s},\mathbf C_p)}_
{\phi\hp{(\mathbf B_{2s},}\mathbf C_p'}D_{A'}^{A_{2s}}
\mathcal Q_{\mathbf A_{2s}}=0,\qquad
\overline\partial^{\hp{\phi}(\mathbf B_{2s}',\mathbf C_p')}_
{\phi\hp{(\mathbf B_{2s}',}\mathbf C_p}D_{A'}^{A_{2s}}
\mathcal Q_{\mathbf A_{2s}}=0
\end{equation}
on $\R$. By virtue of the commutation formulas
\eqref{E:CommutationFormula},
the above equations with $p=r+1$ show
that
\begin{equation}
\label{E:HighestOrderEquations}
\partial^{\hp{\phi}(\mathbf B_{2s},\mathbf C_r}_
{\phi\hp{(\mathbf B_{2s},}
\mathbf C_r'}\mathcal Q_{\mathbf A_{2s-1}}^{A_{2s})}=0,\qquad
\overline \partial^{\hp{\phi}(\mathbf B_{2s}',\mathbf C_r')}_
{\phi\hp{(\mathbf B_{2s}',}(\mathbf C_r}
\mathcal Q_{A_{2s})\mathbf A_{2s-1}}=0
\end{equation}
identically on $\J{E_s}$. Equations
\eqref{E:HighestOrderEquations},
in turn, combined with
\eqref{E:MaySymmetrizeDerivatives},
imply that
\begin{align}
\label{E:HighestOrderTermsReduced}
\partial^{\hp{\phi}\mathbf B_{2s},\mathbf C_r}_
{\phi\hp{(\mathbf B_{2s}}\mathbf C_r'}\mathcal Q_{\mathbf A_{2s}}&=
\e_{A_1}{}^{(B_1}\e_{A_2}{}^{B_2}\cdots\e_{A_{2s}}{}^{B_{2s}}
\mathcal S^{\mathbf C_r)}_{\mathbf C_r'},\nonumber\\
\overline \partial^{\hp{\phi}\mathbf B_{2s}',\mathbf C_r'}_
{\phi\hp{(\mathbf B_{2s}'}\mathbf C_r}\mathcal Q^{\mathbf A_{2s}}&=
\e_{(C_{r-2s+1}}{}^{A_{1}}\e_{C_{r-2s+2}}{}^{A_{2}}\cdots\e_{C_{r}}{}^{A_{2s}}
\mathcal T_{\mathbf C_{r-2s})}^{\mathbf B_{2s}'\mathbf C_{r}'},
\end{align}
for some spinor valued functions
$\mathcal S^{\mathbf C_r}_{\mathbf C_r'}$,
$\mathcal T_{\mathbf C_{r-2s}}^{\mathbf C_{r+2s'}}$
on $\J{E_s}$ symmetric in their indices. Note in
particular that
$\mathcal T_{\mathbf C_{r-2s}}^{\mathbf C_{r+2s'}}$
vanishes if $r<2s$.

Next use equations \eqref{E:SymmetrizedDerivatives} with
$p=r$ together with the commutation formulas
\eqref{E:CommutationFormula} to conclude that
\begin{equation}
\label{E:NextToHighestOrderTerms}
D_{(A'}^{A_{2s}}\partial^{\hp{\phi}\mathbf B_{2s},\mathbf C_r}_
{\phi\hp{\mathbf B_{2s},}\mathbf C_r')}\mathcal Q_{\mathbf A_{2s}} = 0,\qquad
D^{A_{2s}(A'}\overline \partial^{\hp{\phi}\mathbf B_{2s}',\mathbf C_r')}_
{\phi\hp{\mathbf B_{2s},}\mathbf C_r}\mathcal Q_{\mathbf A_{2s}} = 0
\qquad\mbox{on \ \ $\R$}.
\end{equation}
Now substitute expressions
\eqref{E:HighestOrderTermsReduced}
into \eqref{E:NextToHighestOrderTerms} to deduce that
\begin{displaymath}
D^{(C_{r+1}}_{(C_{r+1}'}
\mathcal S^{\mathbf C_r)}_{\mathbf C_r')}=0,\qquad
D_{(C_{r-2s+1}}^{(C_{r+2s+1}'}
\mathcal T_{\mathbf C_{r-2s})}^{\mathbf C_{r+2s}')}=0
\qquad\text{on \ \ $\R$}.
\end{displaymath}
But it is easy to see that the above
equations force
$\mathcal S^{\mathbf C_r}_{\mathbf C_r'}$,
$\mathcal T_{\mathbf C_{r-2s}}^{\mathbf C_{r+2s}'}$
to be independent of the symmetrized derivative variables
$\phi_{\mathbf B_{p+2s}}^{\mathbf B_p'}$,
$p\geq0$, and, consequently, they must
satisfy the Killing spinor equations
\begin{displaymath}
\partial^{(C_{r+1}}_{(C_{r+1}'}
\mathcal S^{\mathbf C_r)}_{\mathbf C_r')}=0,\qquad
\partial_{(C_{r-2s+1}}^{(C_{r+2s+1}'}
\mathcal T_{\mathbf C_{r-2s})}^{\mathbf C_{r+2s}')}=0.
\end{displaymath}
Thus
\begin{displaymath}
\mathcal Q_{\mathbf A_{2s}} =\mathcal S_{\mathbf B_r'}
^{\mathbf B_r}\phi_{\mathbf A_{2s} \mathbf B_r}
^{\hp{\mathbf A_{2s}}\mathbf B_r'} +
\mathcal T_{\mathbf B_{r-2s}}^{\mathbf B_{r+2s}'}
\phiconj_{\mathbf A_{2s}\mathbf B_{r+2s}'}
^{\hp{\mathbf A_{2s}}\mathbf B_{r-2s}} +
\mathcal U_{\mathbf A_{2s}},
\end{displaymath}
where $\mathcal U_{\mathbf A_{2s}}$ only involves
the derivative variables
$\phi_{\mathbf A_{2s+p}}^{\mathbf A_{p}'}$
up to order $r-1$.

Now by the factorization property of Lemma
\ref{L:KillingSpinor}, the Killing spinors
$\mathcal S_{\mathbf B_r'}^{\mathbf B_r}$,
$\mathcal T_{\mathbf B_{r-2s}}^{\mathbf B_{r+2s}}$
can be expressed as a sum of symmetrized products
of $r$ Killing spinors of type $(1,1)$, and
as a sum of symmetrized products of
a Killing spinor of type $(0,4s)$ and
$r-2s$ Killing spinors of type $(1,1)$,
respectively. Thus by Proposition
\ref{P:LeadingOrderTerms}
there is a linear combination
$\mathcal V_{r}$
of the basic symmetries
\begin{displaymath}
Z[\xi;\zeta_1,\dots,\zeta_{r-1}],\qquad
Z[\im\xi;\zeta_1,\dots,\zeta_{r-1}],\qquad
W[\pi;\zeta_1,\dots,\zeta_{r-2s}]
\end{displaymath}
so that on $\R$, the highest order terms in
$\mathcal V_r$
agree with those in $\mathcal Q$,
and, consequently, the symmetry $\mathcal Q$
is equivalent to a linear combination of
the basic symmetries
\eqref{E:HigherConformalSymmetry}--\eqref{E:HigherChiralSymmetry}
with $p=r-1$, $q=r-2s$ and an evolutionary
symmetry of order $r-1$.

Now proceed inductively in the order of
the symmetry.
In the last step the symmetry $\mathcal Q$
is equivalent to a linear combination
of the symmetries
\eqref{E:HigherConformalSymmetry}--\eqref{E:HigherChiralSymmetry}
with $p\leq r-1$, $q\leq r-2s$, and an
evolutionary symmetry $\mathcal V_{o}$
of order $0$. But it is straightforward
to solve the determining equations
\eqref{E:EvolutionarySymmetry} for
$\mathcal V_{o}$ to see that
\begin{displaymath}
\mathcal V_{o,\mathbf A_{2s}}=
a\phi_{\mathbf A_{2s}}+\varphi_{\mathbf A_{2s}},
\end{displaymath}
where $a\in\mathbf C$ is a constant and
$\varphi_{\mathbf A_{2s}} =\varphi_{\mathbf A_{2s}}(x^i)$
is a solution of
the massless free f\/ield equations of spin $s$.
Thus \eqref{E:RemoveTrivialSymmetry} holds.

Finally, the above arguments show that the
vector space of equivalence classes of
symmetries of order $r\geq1$ modulo
symmetries of order $r-1$ is isomorphic
with the real vector space of Killing
spinors of type $(r,r)$, if $r<2s$ and with
the direct sum of the real vector spaces of Killing
spinors of type  $(r,r)$ and $(r-2s,r+2s)$ if $r\geq2s$.
The dimension of the space spanned by the spinorial
symmetries \eqref{E:NonTrivialSymmetries} now
can be computed by adding up the dimensions
given in Lemma~\ref{L:KillingSpinor}.
This concludes the proof of the Theorem.
\end{proof}

\section{Symmetries of Maxwell's equations}
\label{S:MaxwellSymmetries}

In this section we transcribe the spinorial symmetries
of Theorem \ref{T:MainTheorem} for $s=1$ to tensorial
form in order to classify generalized symmetries of Maxwell's
equations
\begin{equation}
\label{E:Maxwell'sEquations}
F_{ij,}{}^j = 0,\qquad *F_{ij,}{}^j=0
\end{equation}
on Minkowski space. Here $F_{ij}=-F_{ji}$
are the components of the
electromagnetic f\/ield tensor $F$
and $*$ stands for the Hodge dual.

We write $\Lambda^2(T^*\mathbf M)\to\mathbf M$
for the associated bundle with coordinates
$F_{ij}$, $i<j$. Then $J^\infty \Lambda^2(T^*\mathbf M)$
is the coordinate bundle{\samepage
\[
\{(x^i,F_{ij},F_{ij,k_1},F_{ij,k_1k_2},\dots)\}\to\{(x^i)\}.
\]
For notational convenience,
we write $F_{ij,k_1\cdots k_p} = -F_{ji,k_1\cdots k_p}$
for $i\geq j$.}

In spinor form the electromagnetic
f\/ield tensor $F$ becomes
\begin{displaymath}
\sigma_{AA'}^i\sigma_{BB'}^j F_{ij} =
\e_{A'B'}\phi_{AB}+\e_{AB}\overline\phi_{A'B'},
\end{displaymath}
while Maxwell's equations \eqref{E:Maxwell'sEquations}
correspond to the spin $s=1$ massless free f\/ield equations
\eqref{E:MasslessEquations} for the electromagnetic
spinor $\phi_{AB} = \phi_{(AB)}$.

A generalized symmetry of Maxwell's equations
in evolutionary form is a vector f\/ield
$
Y = Q_{ij}\partial^{ij}_F
$
satisfying
\begin{equation}
\label{E:MaxwellSymmetryEquations}
D^jQ_{ij}=0,\qquad D^j\mkern -3mu*\mkern-2muQ_{ij}=0
\end{equation}
on solutions of \eqref{E:Maxwell'sEquations}.
If one def\/ines $\mathcal Q_{AB}$ by
\begin{equation}
\label{E:SpinorRepresentativeOfSymmetry}
\sigma_{AA'}^i\sigma_{BB'}^j Q_{ij}=
\e_{A'B'}\mathcal Q_{AB} +
\e_{AB}\overline{\mathcal Q}_{A'B'},
\end{equation}
then it easily follows from
\eqref{E:MaxwellSymmetryEquations},
\eqref{E:SpinorRepresentativeOfSymmetry}
that $\mathcal Q_{AB}$ are the components
of a generalized symmetry of the massless
f\/ield equations of spin $s=1$. Thus, by employing
the correspondence
\eqref{E:SpinorRepresentativeOfSymmetry},
we can obtain
a complete classif\/ication of generalized
symmetries of Maxwell's equations from
the classif\/ication result in Theorem~\ref{T:MainTheorem}.

Symmetries \eqref{E:ScalingSymmetries},
\eqref{E:TrivialSymmetries} clearly
correspond to the symmetries
\begin{equation*}
\mathcal S = F_{ij}\partial^{ij}_F,\qquad
\widetilde{\mathcal S} = *\mkern-1mu F_{ij}\partial^{ij}_F,\qquad
\mathcal E(\mathbf F) = \mathrm F_{ij}\partial^{ij}_F
\end{equation*}
of Maxwell's equations, where $\mathbf F$ is any solution
of \eqref{E:Maxwell'sEquations} with components
$\mathrm F_{ij}=\mathrm F_{ij}(x^k)$.
Conformal symmetries \eqref{E:ConformalSymmetry}
and their duals \eqref{E:DualOfConformalSymmetry}
in turn give rise to the symmetries
\begin{equation}
\label{E:ConformalSymmetries}
\mathcal Z[F;\xi]=\mathcal Z_{ij}[F;\xi]\partial^{ij}_F,\qquad
\mathcal Z[*\mkern-1mu F;\xi]=\mathcal Z_{ij}[*\mkern-1mu F;\xi]\partial^{ij}_F,
\end{equation}
with components
\begin{displaymath}
\mathcal Z_{ij}[F;\xi] =
\xi^kF_{ij,k}-2\partial_{[i}\xi^k{}F_{j]k},
\end{displaymath}
where $\xi$ is a conformal Killing vector on
$\mathbf M$ and where square brackets indicate
skew-sym\-met\-ri\-za\-tion in the enclosed indices.

In order to transcribe the second order chiral
symmetries $\mathcal W[\pi]$ introduced in
\eqref{E:ChiralSymmetry} to symmetries of
Maxwell's equations in physical form,
we f\/irst introduce the following polynomial
tensors on $\mathbf M$.
Let
\begin{align}
\label{E:PolynomialOfOrder0}
p^0_{ijkl} ={} &a^0_{ijkl},\\
\label{E:PolynomialOfOrder1}
p^1_{ijkl}={} &x^{}_{[i}a^1_{j]kl}+x^{}_{[k}a^1_{l]ij}+
(\eta^{}_{[i|[k}a^1_{l]|j]n}+\eta^{}_{[k|[i}a^1_{j]|l]n})x^n,\\
\label{E:PolynomialOfOrder2}
p^2_{ijkl} ={} &a^2_{[i|[k} x^{}_{l]|} x^{}_{j]}
-\dfrac{1}{2}\eta^{}_{[i|[k} a^2_{l]|j]} x^{}_mx^m
\nonumber\\
&{}+\dfrac{1}{2}(\eta^{}_{[i|[k} a^2_{l]n|} x^{}_{j]} +
\eta^{}_{[k|[i} a^2_{j]n|} x^{}_{l]} ) x^n
-\dfrac{1}{6} \eta^{}_{i[k} \eta^{}_{l]j} a^2_{mn} x^m x^n,\\
\label{E:PolynomialOfOrder3}
p^3_{ijkl} ={}&
( x^{}_{[i} a^3_{j]n[k} x^{}_{l]}+ x^{}_{[k} a^3_{l]n[i} x^{}_{j]} ) x^n
+\dfrac{1}{4}(x^{}_{[i} a^3_{j]kl} + x^{}_{[k} a^3_{l]ij} ) x^{}_nx^n
\nonumber\\
&{}+\dfrac{1}{2}( a^3_{mn[i} \eta^{}_{j][k} x^{}_{l]}
+a^3_{mn[k} \eta^{}_{l][i} x^{}_{j]} ) x^m x^n
+\dfrac{1}{4}(a^3_{[i|m[k}\eta^{}_{l]|j]}+
a^3_{[k|m[i}\eta^{}_{j]|l]})x^mx^{}_nx^n,\\
\label{E:PolynomialOfOrder4}
p^4_{ijkl} ={}& (a^4_{m[i|n[k}x^{}_{l]|}x^{}_{j]}-
           \dfrac{1}{2}a^4_{m[i|n[k}\eta^{}_{l]|j]}x^{}_px^p)x^mx^n
          -\dfrac{1}{16}a^4_{ijkl}x^{}_mx^mx^{}_nx^n,
\end{align}
where $a^0_{ijkl}$, $a^1_{ijk}$, $a^2_{ij}$, $a^3_{ijk}$,
$a^4_{ijkl}$ are real constants satisfying{\samepage
\begin{alignat}{5}
\label{E:CoefficientsOfOrder0,4}
& a^{\mathtt{h}}_{ijkl} = a^\mathtt{h}_{[kl][ij]},\qquad &&
a^\mathtt{h}_{[ijkl]} = 0,\qquad && a^\mathtt{h}_{ijk}{}^j = 0,\qquad &&
\mathtt{h}=0,4, & \\
\label{E:CoefficientsOfOrder1,3}
& a^\mathtt{h}_{ijk} = a^\mathtt{h}_{i[jk]},\qquad &&
a^\mathtt{h}_{[ijk]} = 0,\qquad&&
a^\mathtt{h}_{ji}{}^j = 0,\qquad&&
\mathtt{h}=1,3,&\\
\label{E:CoefficientsOfOrder2}
& a^2_{ij} = a^2_{(ij)},\qquad&&
a^2_{i}{}^i = 0, &&&& &
\end{alignat}
and where we raise indices using the inverse $\eta^{ij}$
of the Minkowski metric.}

Note that $a^\mathtt{h}_{ijkl}$, $\mathtt{h}=0,4$, possesses the
symmetries of Weyl's conformal curvature tensor.
Hence their spinor representatives can be written
in the form
\begin{equation*}
a^\mathtt{h}_{II'JJ'KK'LL'} = \e_{IJ}\e_{KL}\alpha^\mathtt{h}_{I'J'K'L'}+
	     \e_{I'J'}\e_{K'L'}\overline\alpha^\mathtt{h}_{IJKL},
	  \qquad \mathtt{h}=0,4,
\end{equation*}
where $\alpha^\mathtt{h}_{I'J'K'L'}$  is a constant,
symmetric spinor.
See, for example,~\cite{Penrose84}. Moreover, it is
easy to verify that by \eqref{E:CoefficientsOfOrder1,3},
\eqref{E:CoefficientsOfOrder2}, the
spinor representatives of $a^2_{ij}$ and
$a^\mathtt{h}_{ijk}$, $\mathtt{h}=1,3$, are given by
\begin{align*}
& a^2_{II'JJ'} = \alpha^2_{IJI'J'}\qquad\text{and}\\
& a^\mathtt{h}_{II'JJ'KK'}=
\e_{JK}\alpha^\mathtt{h}_{II'J'K'}+\e_{J'K'}\overline\alpha^\mathtt{h}_{I'IJK},
\qquad \mathtt{h}=1,3,
\end{align*}
where
$\alpha^1_{II'J'K'}$,
$\alpha^2_{IJI'J'}=\overline\alpha^2_{IJI'J'}$,
$\alpha^3_{II'J'K'}$,
are symmetric constant spinors.

For $0\leq \mathtt{h}\leq4$, let $\mathcal W[F;p^\mathtt{h}]$
denote the evolutionary vector f\/ield with
components
\begin{equation}
\mathcal W_{ij}[F;p^\mathtt{h}] = p^\mathtt{h}_{klm[i}F^{kl}{}_,{}^m{}_{j]}
+\partial_{[i}p^\mathtt{h}_{j]mkl}F^{kl}{}_,{}^m+
\dfrac{3}{5}\partial^mp^\mathtt{h}_{klm[i}F^{kl}{}_{,j]}+
\dfrac{3}{5}\partial^m\partial_{[i}p^\mathtt{h}_{j]mkl}F^{kl},
\label{E:MaxwellChiralSymmetry}
\end{equation}
where $p^\mathtt{h}_{ijkl}$ are the polynomials in
\eqref{E:PolynomialOfOrder0}--\eqref{E:PolynomialOfOrder4},
and write
\begin{align*}
& \mathcal Z[F;\xi,\zeta_1,\dots,\zeta_q] =
\pr \mathcal Z[F;\zeta_1]\cdots
\pr \mathcal Z[F;\zeta_q]Z[F;\xi],\\
& \mathcal W[F;p^\mathtt{h};\zeta_1,\dots,\zeta_q] =
\pr \mathcal Z[F;\zeta_1]\cdots
\pr \mathcal Z[F;\zeta_q]W[F;p^\mathtt{h}]
\end{align*}
for the componentwise Lie derivatives,
where $\mathcal Z[F;\zeta_i]$ stands for the conformal
symmetry
\eqref{E:ConformalSymmetries} associated with the
conformal Killing vector $\zeta_i$.

\begin{lemma}
\label{L:SpinorForms}
Let $p^{\mathtt{h}}_{ijkl}$, $0\leq \mathtt{h}\leq4$,
be the polynomials \eqref{E:PolynomialOfOrder0}--\eqref{E:PolynomialOfOrder4}. Then the spinor
representative of $p^\mathtt{h}_{ijkl}$ is of the form
\begin{equation}
\label{E:SpinorFormOfPolynomials}
p^\mathtt{h}_{II'JJ'KK'LL'} = \e_{IJ}\e_{KL}\pi^\mathtt{h}_{I'J'K'L'}+
\e_{I'J'}\e_{K'L'}\overline\pi^\mathtt{h}_{IJKL},
\end{equation}
where
\begin{alignat}{3}
& \pi^0_{I'J'K'L'} = \alpha^0_{I'J'K'L'},\qquad & &
\pi^1_{I'J'K'L'} = \alpha^1_{L(I'J'K'}x_{L')}^{L}, &
\nonumber
\\
& \pi^2_{I'J'K'L'} = \dfrac{1}{4}\alpha^2_{KL(I'J'}x^{K}_{K'}x^{L}_{L')},\qquad&&
\pi^3_{I'J'K'L'} = -\dfrac{1}{2}\overline
\alpha^3_{JKL(I'}x^{J}_{J'}x^{K}_{K'}x^{L}_{L')},
\label{E:PolynomialsInSpinorForm}&
\\
& \pi^4_{I'J'K'L'}  = \dfrac{1}{4}\overline
\alpha^4_{IJKL}x^{I}_{I'}x^{J}_{J'}x^{K}_{K'}x^{L}_{L'}.\qquad &&&
\nonumber
\end{alignat}
Moreover, on the solution manifold $\mathcal R^\infty(E_1)$,
the spinor representatives of the evolutionary
vector fields $\mathcal W[F;p^\mathtt{h}]$ in
\eqref{E:MaxwellChiralSymmetry}
reduce to
\begin{equation}
\label{E:ChiralSymmetriesInSpinorForm}
\mathcal W_{II'JJ'}[F;p^\mathtt{h}] =
\e_{I'J'}\mathcal W_{IJ}[\pi^\mathtt{h}]+
\e_{IJ}\overline{\mathcal W}_{I'J'}[\pi^\mathtt{h}],
\end{equation}
where $\mathcal W_{IJ}[\pi^\mathtt{h}]$ are the components \eqref{E:ChiralSymmetryComponents} of chiral symmetries
for spin $s=1$ fields.
Thus $W_{ij}[F;p^\mathtt{h}]$, $0\leq \mathtt{h}\leq4$, is a symmetry of Maxwell's
equations. Moreover,
\begin{equation}
\label{E:ChiralPropertyOfW}
\hodge W_{ij}[F;p^\mathtt{h}] = -W_{ij}[*\mkern-1mu F;p^\mathtt{h}].
\end{equation}
\end{lemma}

\begin{proof} The proofs of the equations in the Lemma are based
on straightforward albeit lengthy computations. We will therefore,
as an example, derive equation~\eqref{E:SpinorFormOfPolynomials}
for $\mathtt{h}=2$ and equation~\eqref{E:ChiralSymmetriesInSpinorForm},
and omit the
proofs of the remaining equations in the Lemma.

One can check by a direct computation that the polynomials
$p^\mathtt{h}_{ijkl}$  possess the symmetries of Weyl's conformal
curvature tensor, that is,
\begin{displaymath}
p^\mathtt{h}_{ijkl} = p^\mathtt{h}_{[kl][ij]},\qquad
p^\mathtt{h}_{[ijkl]} = 0,\qquad p^\mathtt{h}_{ijk}{}^j = 0,\qquad
0\leq \mathtt{h}\leq4.
\end{displaymath}
Consequently, we can write the spinor representatives of
$p^\mathtt{h}_{ijkl}$ in the form \eqref{E:SpinorFormOfPolynomials};
see~\cite{Penrose84}.
Then, in particular,
\begin{equation}
\label{E:CoefficientsInSpinorFormSimplified}
\pi^\mathtt{h}_{I'J'K'L'} = \dfrac{1}{4}
p^\mathtt{h}_{P(I'}{}^P{}_{J'|Q|K'}{}^Q{}_{L')}.
\end{equation}
Thus we can f\/ind $\pi^\mathtt{h}_{I'J'K'L'}$ by substituting the spinor
representative of the expression on the right-hand side
of equations
\eqref{E:PolynomialOfOrder0}--\eqref{E:PolynomialOfOrder4}
into
\eqref{E:CoefficientsInSpinorFormSimplified}
and, in each of the resulting terms, symmetrizing
over primed indices. The computations can be further
simplif\/ied by using the observation that given a
tensor $f_{ijkl}$ with the spinor representative
$f_{II'JJ'KK'LL'}$, then the spinor representative
$h_{II'JJ'KK'LL'}$ of $h_{ijkl} = f_{[ij][kl]}$
satisf\/ies
\begin{equation}
\label{E:Shortcut}
h_{PI'}{}^P{}_{J'QK'}{}^Q{}_{L'}=
f_{P(I'J')}{}^P{}_{Q(K'L')}{}^Q.
\end{equation}

Now insert the spinor representative of the right-hand
side of the equation \eqref{E:PolynomialOfOrder2} into
\eqref{E:CoefficientsInSpinorFormSimplified}.
Then by virtue of \eqref{E:Shortcut}, the f\/irst term
yields the expression
\begin{displaymath}
\dfrac{1}{4}\alpha^2_{KL(I'J'}x_{K'}^Kx_{L')}^L.
\end{displaymath}
The spinor forms of each of the
remaining terms on the right-hand
side of \eqref{E:PolynomialOfOrder2}
contain an instance of the
conjugate of the epsilon tensor,
and, consequently, upon symmetrization
over primed indices these terms vanish.
Hence we have that
\begin{displaymath}
\pi^2_{I'J'K'L'} = \dfrac{1}{4}\alpha_{KL(I'J'}x_{K'}^Kx_{L')}^L,
\end{displaymath}
as required.

In order to derive \eqref{E:ChiralSymmetriesInSpinorForm},
we f\/irst write
\begin{equation*}
\fminus_{kl} = \dfrac{1}{2}(F_{kl}-\im\hodge F_{kl}),\qquad
\Wplus_{ij}[F,p^\mathtt{h}] = \dfrac{1}{2}( W_{ij}[F,p^\mathtt{h}]-\im\mspace{1mu} W_{ij}[\hodge F,p^\mathtt{h}]),
\end{equation*}
so that
\begin{gather}
\Wplus_{ij}[F,p^\mathtt{h}] =
      p^\mathtt{h}_{klm[i}\fminus^{kl}{}_,{}^m{}_{j]}
     +\partial_{[i}p^\mathtt{h}_{j]mkl}\fminus^{kl}{}_,{}^m \nonumber\\
\phantom{\Wplus_{ij}[F,p^\mathtt{h}] =}{}
    +\dfrac{3}{5}\partial^mp^\mathtt{h}_{klm[i}\fminus{}^{kl}{}_{,j]} +
     \dfrac{3}{5}\partial^m\partial_{[i}p^\mathtt{h}_{j]mkl}\fminus^{kl} . \label{E:SelfDualW}
\end{gather}
We also have that
$
\fminus_{KK'LL'} = \e_{KL}\phiconj_{K'L'}
$.

Note that by virtue of the Killing spinor equations
\begin{displaymath}
\partial_{I'}^P\pi^\mathtt{h}_{J'K'L'M'} =
-\dfrac{4}{5}\e_{I'(J'}\partial^{PP'}\pi^\mathtt{h}_{K'L'M')P'}.
\end{displaymath}
Hence
\begin{equation}
\label{E:IJSymmDerOfPi}
\partial_{P(I'}\pi^\mathtt{h}_{J')K'L'M'}\phiconj^{K'L'M'P}=
\dfrac{3}{5}\partial_P^{P'}\pi^\mathtt{h}_{K'L'P'(I'}\phiconj_{J')}{}^{K'L'P}.
\end{equation}
Consider, for example, the spinor representative of the second
term on the right-hand side of equation \eqref{E:SelfDualW}.
On the solution manifold $\mathcal R^\infty(E_1)$ we have
\begin{align}
\label{E:SecondTermInSelfDualW}
\sigma^i_{II'}\sigma^j_{JJ'}
&\partial_{[i}p^\mathtt{h}_{j]mkl}\fminus^{kl}{}_,{}^m
\nonumber\\
&=-\partial_{II'}\pi^\mathtt{h}_{J'K'L'M'}\phiconj_{J}{}^{K'L'M'}+
\partial_{JJ'}\pi^\mathtt{h}_{I'K'L'M'}\phiconj_{I}{}^{K'L'M'}
\nonumber\\
&=
\e_{I'J'}\partial_{P'(I}\pi^{\mathtt{h}K'L'M'P'}\phiconj_{J)K'L'M'}
-\e_{IJ}\partial_{P(I'}\pi^\mathtt{h}_{J')K'L'M'}\phiconj^{K'L'M'P}\\
&=\e_{I'J'}\partial_{P'(I}\pi^{\mathtt{h}K'L'M'P'}\phiconj_{J)K'L'M'}
-\dfrac{3}{5}\e_{IJ}
\partial_P^{P'}\pi^\mathtt{h}_{K'L'P'(I'}\phiconj_{J')}{}^{K'L'P},
\nonumber\end{align}
where we used \eqref{E:IJSymmDerOfPi}.
Similarly, on $\mathcal R^\infty(E_1)$,
we see that
\begin{gather}
\label{E:FirstTermInSelfDualW}
\sigma^i_{II'}\sigma^j_{JJ'}p^\mathtt{h}_{klm[i}\fminus^{kl}{}_,{}^m{}_{j]}
=\e_{I'J'}\pi^\mathtt{h}_{K'L'M'P'}\phiconj_{IJ}{}^{K'L'M'P'},\qquad\mbox{and}\\
\label{E:ThirdTermInSelfDualW}
\sigma^i_{II'}\sigma^j_{JJ'}\partial^m
p^\mathtt{h}_{klm[i}\fminus^{kl}{}_,{}_{j]}
=\e_{I'J'}\partial_{P'(I}\pi^{\mathtt{h}K'L'M'P'}\phiconj_{J)K'L'M'}\\
\phantom{\sigma^i_{II'}\sigma^j_{JJ'}\partial^m
p^\mathtt{h}_{klm[i}\fminus^{kl}{}_,{}_{j]}=}{} -\e_{IJ}\partial_{PP'}\pi^{\mathtt{h}K'L'P'}{}_{(I'}\phiconj_{J')K'L'}{}^P.\nonumber
\end{gather}
Moreover, by virtue of \eqref{E:DlambertOnPi},
\begin{align}
\sigma^i_{II'}\sigma^j_{JJ'}&\partial^m
\partial_{[i}p^\mathtt{h}_{j]mkl}\fminus^{kl}\nonumber\\
&=\e_{I'J'}\partial_{M'I}\partial_{JP'}\pi^{\mathtt{h}K'L'M'P'}\phiconj_{K'L'}-
\e_{IJ}\partial_{PM'}\partial^{P}_{(I'}\pi^\mathtt{h}_{J')}{}^{K'L'}{}^{M'}\phiconj_{K'L'}
\label{E:LastTermInSelfDualW}\\
&=\e_{I'J'}\partial_{M'I}\partial_{JP'}\pi^{\mathtt{h}K'L'M'P'}\phiconj_{K'L'}
\nonumber\end{align}
on $\mathcal R^\infty(E_1)$.

Now equations \eqref{E:SecondTermInSelfDualW}--\eqref{E:LastTermInSelfDualW}
together show that the spinor form of
$\Wplus_{ij}[F,p^\mathtt{h}]$ is
\begin{align}
\Wplus_{II'JJ'}[F,p^\mathtt{h}] ={}&\e_{I'J'}\big(\pi^{\mathtt{h}K'L'M'P'}\phiconj_{IJK'L'M'P'}
\nonumber\\
&{}+\dfrac{8}{5}\partial_{P'(I}\pi^{\mathtt{h}K'L'M'P'}\phiconj_{J)K'L'M'}
+\dfrac{3}{5}\partial_{M'I}\partial_{JP'}\pi^{\mathtt{h}K'L'M'P'}\phiconj_{K'L'}\big)
\nonumber\\
={}&\e_{I'J'}\mathcal W_{IJ}[\pi^\mathtt{h}],
\nonumber
\end{align}
which immediately yields equation
\eqref{E:ChiralSymmetriesInSpinorForm}.
\end{proof}

\begin{theorem}
\label{T:MaxwellSymmetries}
The space of equivalence classes of
symmetries of Maxwell's equations of order $r$,
$r\geq2$, is spanned by the symmetries
\begin{alignat*}{3}
&\mathcal E({\mathbf F}),\quad\mathcal S,
\quad\widetilde{\mathcal S},&&&&\nonumber \\
&\mathcal Z[F;\xi,\zeta_1,\dots,\zeta_q],&\qquad
&\mathcal Z[*\mkern-1mu F;\xi,\zeta_1,\dots,\zeta_q],&\qquad &q\leq r-1,\\
&\mathcal W[F;p^\mathtt{h};\zeta_1\dots,\zeta_q],&\qquad
&\mathcal W[*\mkern-1mu F;p^2;\zeta_1,\dots,\zeta_q],&\qquad
&0\leq \mathtt{h}\leq4,\qquad q\leq r-2,
\nonumber
\end{alignat*}
where $\mathbf F$ is an arbitrary solution of Maxwell's
equations, $\xi$, $\zeta_1$,\dots,$\zeta_r$, are conformal
Killing vectors and $p^\mathtt{h}_{ijkl}$, $0\leq \mathtt{h}\leq4$, are the
polynomials given in
\eqref{E:PolynomialOfOrder0}--\eqref{E:PolynomialOfOrder4}.
Apart from the trivial symmetries
$\mathcal E(\mathbf F)$, there are
\[
d_r=(r+1)(r+3)(r^4+8r^3+17r^2+4r+6)/9
\]
independent symmetries of Maxwell's
equations of order $r\geq2$.
\end{theorem}

\begin{proof} First note that if
$Y = Q_{ij}\partial^{ij}_F$ is a
symmetry of Maxwell's equations in
evolutionary form with the spinor
representative $\mathcal Q$ determined
by \eqref{E:SpinorRepresentativeOfSymmetry}
then the spinor representative of
the symmetry $\pr Z[F;\zeta] Y$
is $\pr Z[\zeta]\mathcal Q$.
Thus, in light of Theorem
\ref{T:MainTheorem} and by linearity,
it suf\/f\/ices to show that
the spinorial symmetries
$\mathcal E(\varphi)$, $\mathcal S$,
$\widetilde{\mathcal S}$,
$\mathcal Z[\xi]$, $\mathcal Z[\im\xi]$,
$\mathcal W[\pi]$, where $\xi$ is a real conformal
Killing vector and $\pi$ is a type $(0,4)$
Killing spinor, are contained in the
span of the spinorial symmetries
corresponding to
\[\mathcal E(\mathbf F),\quad
\mathcal S, \quad \widetilde{\mathcal S},\quad
\mathcal Z[F;\xi],\quad \hodge\mathcal Z[F;\xi],\quad
\hodge\mathcal W[F;p^2],\quad \mathcal W[F;p^\mathtt{h}],\qquad 0\leq \mathtt{h}\leq 4,
\]
under the identif\/ication
\eqref{E:SpinorRepresentativeOfSymmetry}.

Obviously the symmetries
$\mathcal E(\mathbf F)$, $\mathcal S$,
$\widetilde{\mathcal S}$,
$\mathcal Z[F;\xi]$, $*\mathcal Z[F;\xi]$
correspond via
\eqref{E:SpinorRepresentativeOfSymmetry}
to the spinorial symmetries
$\mathcal E(\varphi)$, $\mathcal S$,
$\widetilde{\mathcal S}$
$\mathcal Z[\xi]$, $\mathcal Z[\im\xi]$,
where $\varphi$ is the spinorial counterpart
of the solution $\mathbf F$ of
Maxwell's equations.

Next, by \eqref{E:PolynomialsInSpinorForm},
the spinor f\/ields $\pi^\mathtt{h}_{I'J'K'L'}$
corresponding to the polynomials $p^\mathtt{h}_{ijkl}$
span the
space of Killing spinors of type $(0,4)$
provided that we allow $\alpha^2_{IJI'J'}$
also to take complex values. Thus, in light of
\eqref{E:ChiralSymmetriesInSpinorForm},
the f\/irst part of the Theorem now follows from
the observation that if $\mathcal W[\pi]$ is
the spinorial symmetry corresponding
to $\mathcal W[F;p^\mathtt{h}]$, then
$\mathcal W[\im\pi]$ corresponds to
the symmetry $\widehat{\mathcal W}[F;p^\mathtt{h}]=
-\hodge\mathcal W[F;p^\mathtt{h}]$.

Finally, the dimension count follows from the
dimension count in Theorem \ref{T:MainTheorem}.
This concludes the proof of the Theorem.
\end{proof}

\begin{remark}
The new chiral symmetries $\mathcal W[F;p^\mathtt{h}]$ are
physically interesting since, as evidenced by
equation \eqref{E:ChiralPropertyOfW},
they possess odd parity, i.e., chirality, under
the duality transformation interchanging the electric
and magnetic f\/ields. Hence, in a marked contrast to
the conformal symmetries $\mathcal Z[F;\xi]$,
which behave equivariantly under the duality transformation,
the new second order symmetries, when regarded as
dif\/ferential operators acting on solutions of Maxwell's
equations, map self-dual electromagnetic f\/ields to
anti-self-dual f\/ields and vice versa.
\end{remark}

\section{Applications to the Dirac operator on Minkowski space}
\label{S:Dirac}

Recall that a Dirac spinor consists of a pair
\[
\mathbf{\Psi} = \begin{pmatrix}\psi^{A'}\\ \varphi_{A}\end{pmatrix}
\]
of spinor f\/ields of type $(0,1)$ and $(1,0)$ (see, e.g.,~\cite{Ward90}),
with the conjugate spinor $\mathbf\Psi^*$ given by
\[
\mathbf{\Psi}^* = \begin{pmatrix}\overline{\varphi}^{A'}\\ \overline{\psi}_{A}\end{pmatrix}.
\]

A vector
$v^i\in\mathbf M$ determines a linear transformation
$\vslash$ on Dirac spinors given by
\begin{equation*}
\vslash\mathbf{\Psi} =
\begin{pmatrix}
v^i\sigma_{i}^{AA'}\varphi_A \\
v^i\sigma_{iAA'}\psi^{A'}
\end{pmatrix}.
\end{equation*}
Then the Dirac $\gamma$-matrices are def\/ined
by the condition
\begin{equation*}
v^i\gamma_i\mathbf{\Psi} = \vslash\mathbf{\Psi},
\end{equation*}
so that
\begin{equation*}
\gamma_i = \begin{pmatrix} 0 & \sigma_{i}^{A'B} \\ \sigma_{iAB'} & 0
\end{pmatrix},\qquad 0\leq i\leq 3.
\end{equation*}
Moreover, as is customary, we will write
\[
\gamma_5 = -\im\gamma_0\gamma_1\gamma_2\gamma_3 =
\begin{pmatrix} \Bbb I&\hphantom{-}0\\ 0&-\Bbb I\end{pmatrix}.
\]

The Weyl system, or the
massless Dirac equation, is given by
\begin{equation}\label{E:masslessDirac}
\spartial\mathbf\Psi =0,
\end{equation}
where
\begin{equation*}
\spartial = \gamma^i\dfrac{\partial}{\partial x^i}
\end{equation*}
is the Dirac operator. As an application of the methods
developed in Section~\ref{S:MainResults}
we present here a
complete classif\/ication of generalized symmetries of
the Weyl system on Minkowski space.

Evidently, the Weyl system is equivalent
to the decoupled pair
\begin{equation*}
\varphi_{A,}^{\hphantom{A,}AA'}=0,\qquad
\psi^{A'}{}_{,AA'} = 0
\end{equation*}
of spin $s=\tfrac{1}{2}$ massless free f\/ield equation and its
conjugate equation, and, consequently, we will be able to
analyze symmetries of \eqref{E:masslessDirac}
by the methods employed in the proof of the classif\/ication result in Theorem~\ref{T:MainTheorem}.

Now an evolutionary vector f\/ield takes
the form
\[
X = Q^{A'}\partial_{\psi^{A'}}+
\overline{Q}^A\overline{\partial}_{\psi^A}+
R_{A}\partial_{\varphi_A}+
\overline{R}_{A'}\overline{\partial}_{\varphi_{A'}},
\]
where the characteristic
\[
\mathcal Q =
\begin{pmatrix}Q^{A'}\\ R_{A}\end{pmatrix}
\]
can be assumed to depend on the spacetime variables
$x^i$ and the exact sets of f\/ields, that is, the
symmetrized derivatives
\[
\psi^{A'\mathbf A_{p}'}_{\hphantom{A'}\mathbf A_{p}}=\psi^{(A'\mathbf A_p')}_{\hphantom{(A'},\mathbf A_p},\qquad
\varphi_{A\mathbf A_p}^{\hphantom{A}\mathbf A_p'}=
\varphi_{(A,\mathbf A_p)}^{\hphantom{(A}\mathbf A_p'}
\]
of the components of the Dirac spinor.

As in the case of massless free f\/ields, the Weyl system obviously
admits scaling, conformal and chiral symmetries and their duals.
The scaling symmetry and its dual possess the characteristics
\[
S[\mathbf\Psi] = \mathbf\Psi,\qquad
S[\im\mathbf\Psi] = i\mathbf\Psi.
\]
Let $\xi^{CC'}$ be a real conformal Killing vector,
$\pi^{B'C'}$ a type $(0,2)$ Killing spinor.
We write
\begin{align*}
\mathcal Z[\mathbf\Psi;\xi] &=
\begin{pmatrix}
\xi^{CC'}\psi^{A'}_{CC'} + \dfrac{1}{2}(\partial^{A'}_{C}\xi^{CC'})\psi_{C'}
+\dfrac{1}{8}(\partial_{CC'}\xi^{CC'})\psi^{A'}\\
\noalign{\medskip}
\xi^{CC'}\varphi_{ACC'} + \dfrac{1}{2}(\partial_{AC'}\xi^{CC'})\varphi_C
+\dfrac{1}{8}(\partial_{CC'}\xi^{CC'})\varphi_A
\end{pmatrix},\\
\mathcal W[\mathbf\Psi;\pi] &=
\begin{pmatrix}
\overline{\pi}^{BC}\varphi^{A'}_{BC}
+\dfrac{2}{3}\,\partial^{A'}_{C}\overline{\pi}^{BC}\varphi_{B}\\
\noalign{\medskip}
\pi^{B'C'}\psi_{AB'C'}
+\dfrac{2}{3}\,\partial_{AC'}\pi^{B'C'}\psi_{B'}
\end{pmatrix}
\end{align*}
for the characteristics of the basic conformal and chiral
symmetries of the Weyl system \eqref{E:masslessDirac}.
Moreover, we def\/ine $\mathcal Z[\im\mathbf\Psi;\xi]$
in the obvious fashion, and
denote the componentwise Lie derivatives of
$\mathcal Z[\mathbf\Psi;\xi]$,
$\mathcal Z[\im\mathbf\Psi;\xi]$,
$\mathcal W[\mathbf\Psi;\pi]$ with respect to the
conformal symmetries corresponding to
$\zeta_1$,\dots,$\zeta_p$ by
\[
\mathcal Z[\mathbf\Psi;\xi,\zeta_1,\dots,\zeta_p],\qquad
\mathcal Z[\im\mathbf\Psi;\xi,\zeta_1,\dots,\zeta_p],\qquad
\mathcal W[\mathbf\Psi;\pi;\zeta_1,\dots,\zeta_p].
\]

\begin{theorem}
\label{T:DiracTheorem}
Let ${\mathcal Q}$
be a generalized symmetry of order $r$
of the Weyl system \eqref{E:masslessDirac}
in evolutionary form.
Then ${\mathcal Q}$ is
equivalent to a symmetry $\widehat{\mathcal Q}$ of
order at most $r$ in evolutionary form
which can be written as
\begin{equation*}
\widehat{\mathcal Q} = \mathcal V+\mathcal E[\Psi],
\end{equation*}
where $\mathcal E[\Psi]$ is
an elementary symmetry corresponding to a solution
$\Psi=\Psi(x^i)$
of the Weyl system, and where $\mathcal V$
is equivalent to a symmetry
that is a linear combination of the symmetries
\begin{alignat}{3}
& \mathcal S[\mathbf\Psi],\qquad
\mathcal S[\im\mathbf\Psi],\qquad&&
\gamma_5\mathcal S [\mathbf\Psi],\qquad
\gamma_5\mathcal S[\im\mathbf\Psi],& \nonumber\\
& \mathcal S[\mathbf\Psi^*],\qquad
\mathcal S[\im\mathbf\Psi^*],\qquad&&
\gamma_5\mathcal S [\mathbf\Psi^*],\qquad
\gamma_5\mathcal S[\im\mathbf\Psi^*],& \nonumber\\
& \mathcal Z[\mathbf\Psi;\xi,\zeta_1,\dots,\zeta_p],\qquad&&
\gamma_5\mathcal Z [\mathbf\Psi;\xi,\zeta_1,\dots,\zeta_p],& \nonumber\\
& \mathcal Z[\im\mathbf\Psi;\xi,\zeta_1,\dots,\zeta_p],\qquad&&
\gamma_5\mathcal Z[\im\mathbf\Psi;\xi,\zeta_1,\dots,\zeta_p],\label{E:NonTrivialSymmetriesD}&\\
& \mathcal Z[\mathbf\Psi^*;\xi,\zeta_1,\dots,\zeta_p],\qquad&&
\gamma_5\mathcal Z[\mathbf\Psi^*;\xi,\zeta_1,\dots,\zeta_p],\nonumber& \\
& \mathcal Z[\im\mathbf\Psi^*;\xi,\zeta_1,\dots,\zeta_p],\qquad&&
\gamma_5\mathcal Z[\im\mathbf\Psi^*;\xi,\zeta_1,\dots,\zeta_p],
\nonumber& \\
& \mathcal W[\mathbf\Psi;\pi;\zeta_1,\dots,\zeta_p],\qquad&&
\gamma_5\mathcal W[\mathbf\Psi;\pi;\zeta_1,\dots,\zeta_p],\nonumber&\\
& \mathcal W[\mathbf\Psi^*;\pi;\zeta_1,\dots,\zeta_p],\qquad&&
\gamma_5\mathcal W[\mathbf\Psi^*;\pi;\zeta_1,\dots,\zeta_p],\nonumber&
\end{alignat}
where $p\leq r-1$.
In particular, the dimension $d_r$ of the
real vector space of equivalence classes
of spinorial symmetries of order at most
$r$ spanned by the symmetries
\eqref{E:NonTrivialSymmetriesD} is
\begin{gather*}
d_r = \dfrac{2}{9}\big[(r+1)^2(r+2)^2(r+3)^2
+((r+1)^2-1)((r+2)^2-1)((r+3)^2-1)\big],\quad
\text{for \ $r\geq 1$}.
\end{gather*}
\end{theorem}

\begin{proof} The proof of
Theorem \ref{T:DiracTheorem} follows
very much the same lines as the proof of Theorem
\ref{T:MainTheorem}, and we can safely omit the details.
\end{proof}

\subsection*{Acknowledgements}

The research of the f\/irst author is supported in part by
the NSF Grants DMS 04--53304 and OCE 06--21134,
and that of the second author by an NSERC grant.

\pdfbookmark[1]{References}{ref}
\LastPageEnding


\begin{thebibliography}{99}

\footnotesize\itemsep=0pt




\bibitem{sAjP1}
Anco S.C., Pohjanpelto J.,
Classif\/ication of local conservation laws of Maxwell's equations,
{\it Acta Appl. Math.} {69} (2001), 285--327, \href{http://arxiv.org/abs/math-ph/0108017}{math-ph/0108017}.

\bibitem{sAjP2}
Anco S.C., Pohjanpelto J.,
Conserved currents of massless f\/ields of spin $s\geq 1/2$,
{\it Proc. R. Soc. Lond. A.} {\bf 459} (2003), 1215--1239, \href{http://arxiv.org/abs/math-ph/0202019}{math-ph/0202019}.

\bibitem{sAjP3}
Anco S.C., Pohjanpelto J.,
Symmetries and currents of massless neutrino f\/ields, electromagnetic and graviton f\/ields, in Symmetry in Physics, Editors
P.~Winternitz, J.~Harnad, C.S.~Lam and J.~Patera,
{\it CRM Proceedings and Lecture Notes}, Vol.~34,
AMS, Providence, RI, 2004, 1--12, \href{http://arxiv.org/abs/math-ph/0306072}{math-ph/0306072}.

\bibitem{AnTo96}
Anderson I.M., Torre C.G.,
Classif\/ication of local generalized symmetries
for the vacuum Einstein equations,
{\it Comm. Math. Phys.} {\bf 176} (1996), 479--539, \href{http://arxiv.org/abs/gr-qc/9404030}{gr-qc/9404030}.

\bibitem{BeKr04}
Benn I.M., Kress J.M., First order Dirac symmetry operators,
{\it Classical Quantum Gravity} {\bf 21} (2004), 427--431.

\bibitem{book}
Bluman G., Anco S.C.,
Symmetry and integration methods for dif\/ferential equations,
Springer, New York, 2002.

\bibitem{DuLiVi88}
Durand S.,  Lina J.-M., Vinet L., Symmetries of the massless Dirac
equations in Minkowski space,
{\it Phys. Rev. D} {\bf 38} (1988), 3837--3839.

\bibitem{FuNi87}
Fushchich W.I., Nikitin, A.G.,
On the new invariance algebras and superalgebras
of relativistic wave equations,
{\it J. Phys. A: Math. Gen.} {\bf 20} (1987), 537--549.

\bibitem{Kalnins86}
Kalnins E.G., Miller W. Jr., Williams G.C.,
Matrix operator symmetries of the Dirac equation
and separation of variables,
{\it J. Math. Phys.} {\bf 27} (1986), 1893--1900.

\bibitem{Kalnins92}
Kalnins E.G., McLenaghan R.G., Williams G.C.,
Symmetry operators for Maxwell's
equations on curved space-time,
{\it Proc. R. Soc. Lond. A} {\bf 439} (1992), 103--113.

\bibitem{Kumei75}
Kumei S.,
Invariance transformations, invariance group transformations
and invariance groups of the sine-Gordon equations,
{\it J. Math. Phys.} {\bf 16} (1975), 2461--2468.

\bibitem{MaShWi01} Martina L., Sheftel M.B., Winternitz P.,
{Group foliation and non-invariant solutions of the heavenly equation},
{\it J. Phys. A: Math. Gen.} {\bf 34} (2001), 9243--9263, \href{http://arxiv.org/abs/math-ph/0108004}{math-ph/0108004}.

\bibitem{Mikhailov91}
Mikhailov A.V., Shabat A.B., Sokolov V.V.,
The symmetry approach to classif\/ication
 of integrable equations, in
What Is Integrability?, Editor V.E. Zakharov,
Springer, Berlin, 1991, 115--184.

\bibitem{Miller77}
Miller W. Jr.,
Symmetry and separation of variables,
Addison-Wesley, Reading, Mass., 1977.

\bibitem{Niki91}
Nikitin A.G., A complete set of symmetry operators
for the Dirac equation, {\it Ukrainian Math. J.} {\bf 43} (1991), 1287--1296.


\bibitem{Olver93}
Olver P.J.,
Applications of Lie groups to dif\/ferential equations,
2nd ed.,
Springer, New York, 1993.

\bibitem{Penrose84}
Penrose R., Rindler W.,
Spinors and space-time. Vol.~1: Two-spinor calculus
and relativistic f\/ields, Vol. 2: Spinor and twistor methods
in space-time geometry,
Cambridge University Press, Cambridge, 1984.

\bibitem{Review95}
Pohjanpelto J.,
Symmetries, conservation laws, and Maxwell's equations,
in Advanced Electromagnetism: Foundations, Theory and Applications,
Editors T.W. Barrett and D.M. Grimes,
World Scientif\/ic, Singapore, 1995, 560--589.

\bibitem{Jp02YM}
Pohjanpelto J., Classif\/ication of generalized symmetries
of the Yang--Mills f\/ields with a semi-simple structure group,
{\it Differential Geom. Appl.} {\bf 21} (2004), 147--171, \href{http://arxiv.org/abs/math-ph/0109021}{math-ph/0109021}.

\bibitem{Ward90}
Ward R.S., Wells R.O. Jr.,
Twistor geometry and f\/ield theory,
Cambridge University Press, Cambridge, 1990.

\end{thebibliography}
\end{document}